\documentclass[structabstract]{aa}
\usepackage{amsmath}
\usepackage{graphicx}
\usepackage{txfonts}
\usepackage{natbib}
\bibpunct{(}{)}{;}{a}{}{,} 
\begin{document}
%
%
 \newcommand{\bhyi}{$\beta$~Hyi}

   \title{Asteroseismic modelling of the solar-type subgiant star $\beta$~Hydri}

    \author{I.\,M.\,Brand\~ao$^{\star}$\inst{1,2}, G.\,Do$\rm \breve{g}$an\thanks{I. M. Brand\~ao and G. Do$\rm \breve{g}$an contributed equally to this work.}\inst{3}, J.\,Christensen-Dalsgaard\inst{3}, M.\,S.\,Cunha\inst{1}, T.\,R.\,Bedding\inst{4}, T.\,S.\,Metcalfe\inst{5}, H.\,Kjeldsen\inst{3}, H.\,Bruntt\inst{3,6}, T.\,Arentoft\inst{3}}
   \institute{Centro de Astrof\'isica da Universidade do Porto, Rua das Estrelas, 4150-762 Porto, Portugal\\
              \email{isa@astro.up.pt}
         \and
        Departamento de F\'isica e Astronomia, Faculdade de Ci\^encias da Universidade do Porto, Portugal
    \and
    Department of Physics and Astronomy, Aarhus University, DK-8000 Aarhus C, Denmark
    \and
    Sydney Institute for Astronomy (SIFA), School of Physics, University of Sydney, Australia
    \and
    High Altitude Observatory and Technology Development Division, NCAR, Boulder, Colorado, USA
    \and
    Observatoire de Paris, LESIA, 5 place Jules Janssen, 92195 Meudon Cedex, France
             }

   \date{Received ; accepted }


  \abstract
   {Comparing models and
data of pulsating stars is a powerful way to understand the stellar structure better. Moreover,
such comparisons are necessary to make improvements to the physics of the stellar models, since
they do not yet perfectly represent either the interior or especially the surface layers of
stars. Because $\beta$~Hydri is an evolved solar-type pulsator with mixed modes in its frequency
spectrum, it is very interesting for asteroseismic studies.}
   {The goal of the present work is to search for a
representative model of the solar-type
star $\beta$~Hydri, based on up-to-date non-seismic and
seismic data.}
   {We present a revised list of frequencies for 33 modes, which we
  produced by analysing the power spectrum of the published observations again
  using a new weighting scheme that minimises the daily sidelobes. We ran several grids of evolutionary models with
different input parameters and different physics, using the stellar evolutionary code ASTEC. For
the models that are inside the observed error box of $\beta$~Hydri, we computed their frequencies
with the pulsation code ADIPLS. We used two approaches to find the model that oscillates
with the frequencies that are closest to the observed frequencies of $\beta$~Hydri: (i) we assume
that the best model is the one that reproduces the star's interior based on the radial
oscillation frequencies alone, to which we have applied the correction for the
near-surface effects; (ii) we assume that the best model is the one that produces the lowest value of
the chi-square ($\chi^2$), i.e. that minimises the difference between the
observed frequencies of all available modes and the model predictions,
 after all model frequencies are corrected for near-surface effects.}
   {We show that after applying a correction for near-surface
effects to the frequencies of the best models, we can reproduce the observed modes well,
including those that have mixed mode character. The model that gives the lowest value of the
$\chi^2$ is a post-main-sequence model with a mass of 1.04 M$_{\odot}$ and a metallicity slightly
lower than that of the Sun. Our results underscore the importance of having individual
frequencies to constrain the properties of the stellar model.}
   {}

   \keywords{Asteroseismology - Stars: solar-type - Stars: individual: $\beta$~Hydri}

\authorrunning
{I.\,M.\,Brand\~ao, G.\,Do$\rm \breve{g}$an et al.}
   \maketitle
%

\section{Introduction}
\label{sect:intro} The source $\beta$~Hydri (\bhyi, HD 2151, HR 98, HIP 2021) is a single, bright subgiant
star (m$_{V}$=2.80) that is clearly visible to the naked eye about $12^{\circ}$ from the South
Pole. It is the closest subgiant star, with a spectral and luminosity type between G2\,IV
\citep{hoffleit95,evans57} and G0\,V \citep{gray06}, and it is one of the oldest stars in the
solar Galactic neighbourhood. It is frequently regarded as representing the future of
the Sun \citep{dravins93i,dravins93ii,dravins93iii}, making it a particularly interesting object
of study.

Improvements to the fundamental parameters of \bhyi~have been presented in a number of
recent papers. Recent interferometric measurements of \bhyi~have yielded an accurate (0.8\%)
angular diameter for this star \citep{north07}. Also, the \emph{Hipparcos} parallax of \bhyi~has
been improved from an uncertainty of 0.4\% \citep{esa97} to 0.08\% \citep{van07}. The combination
of these two values gives a direct measure of \bhyi's radius with high accuracy. Moreover,
since the bolometric flux of this star is known \citep{blackwell98}, its position in the
Hertzprung-Russell (HR) diagram is, in principle, well-constrained.

\cite{frandsen87} and \cite{edmonds95} made unsuccessful attempts to
detect stellar oscillations in \bhyi, placing upper limits on the
p-mode amplitudes. \cite{bedding01} and \cite{carrier01} finally
confirmed the presence of solar-like oscillations in \bhyi~and
estimated the large frequency separation $\delta\nu$ to be about
$55\,\mu {\rm Hz}$, but were unable to identify individual mode frequencies. Subsequently,
\cite{bedding07} observed \bhyi~during more than a week with the
high-precision spectrographs HARPS and UCLES. Besides confirming the
oscillations detected in their previous observations in 2000, they
were able to identify 28 oscillation modes that included some mixed modes of spherical degree $l=1$.
Mixed modes occur in stars that
have left the main-sequence stage of their evolution
\citep[e.g.][]{osaki75,aizenman77}, and they provide useful
information about the core. The presence of mixed modes, together
with the availability of very precise non-seismic and seismic data
for \bhyi, places the star in a privileged position for
asteroseismic studies \citep[e.g.][]{cunha07}.

Theoretical models of \bhyi~based on its seismic and non-seismic data have been published by
\cite{fernandes03}, \cite{dimauro03}, and \cite{guln09}. \cite{fernandes03} examined the position
of \bhyi~in the HR diagram by first considering the non-seismic data of the star. In order to
estimate the mass of \bhyi, they used available seismic data, namely the large frequency
separation, to remove partially the helium-content vs mass degeneracy that exists when only
non-seismic observational constraints are used. They also emphasized the usefulness of individual
frequencies to constrain the age of \bhyi~due to the presence of mixed modes in its observed
oscillation spectrum. \cite{dimauro03} computed models of \bhyi, also based on its global
parameters. They used the oscillation frequencies of \bhyi~to compare with the model frequencies.
Their theoretical models reproduced the observed oscillation spectrum of \bhyi well, as well as
the observed large and small frequency separations, after they applied an ad-hoc shift to the
computed frequencies. In fact, when comparing the computed and the observed frequencies, one
should bear in mind that there may be an offset between them. This offset is well known from
helioseismology and is also present when comparing the observed and computed frequencies for
other stars. It arises from improper modelling of the surface layers of stars. \cite{kjeldsen08}
used solar data to derive an empirical correction for the near-surface offset which can be
applied to stars other than the Sun. In our work, we apply this empirical correction to the model
frequencies of \bhyi~before comparing to the observed ones. We extend the analysis of
\cite{guln09}, and also present a detailed discussion on the application of the near-surface
correction.
\begin{figure*}
\begin{center}
$\begin{array}{cc}
\includegraphics[width=9cm,angle=0]{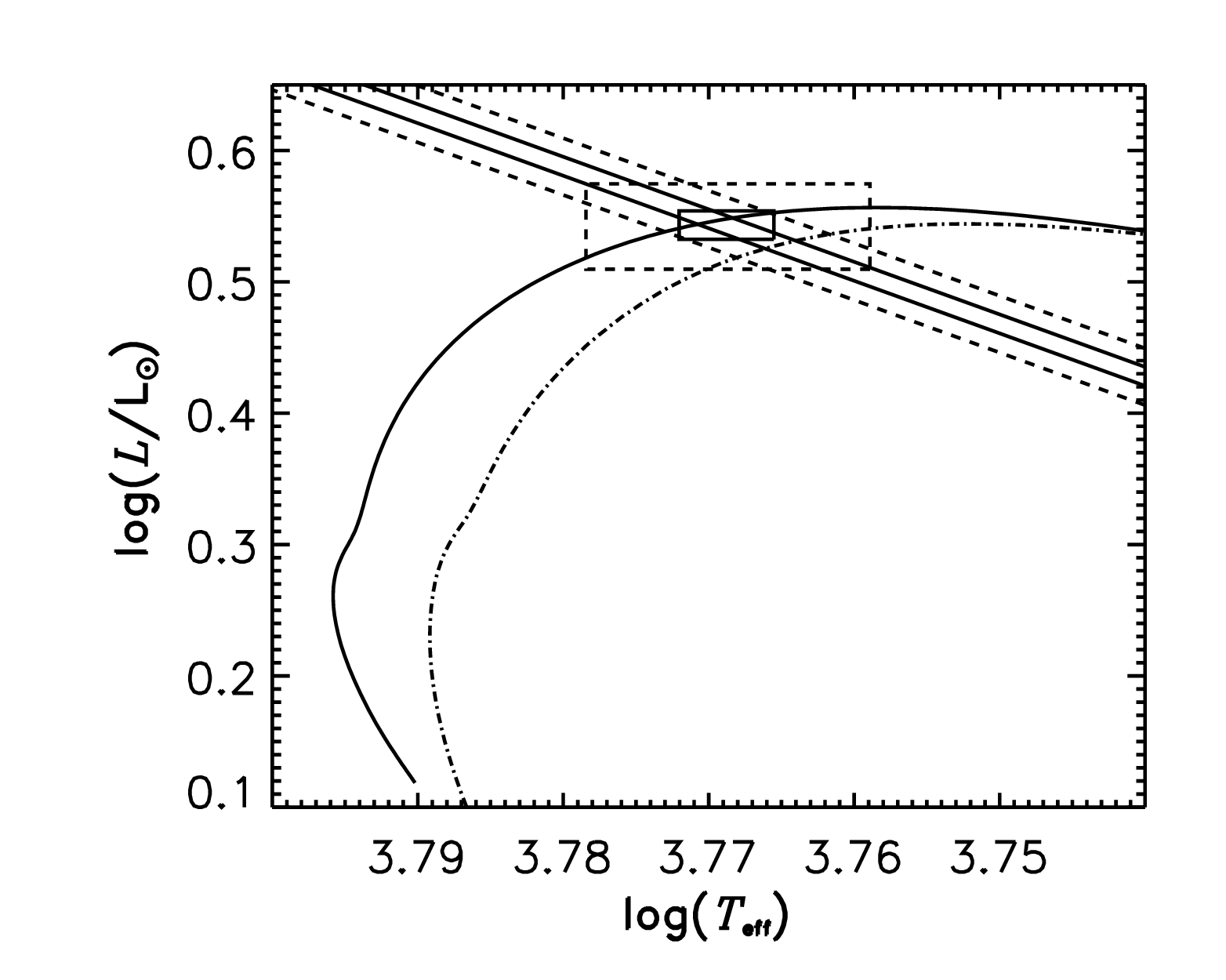} &
\includegraphics[width=9cm,angle=0]{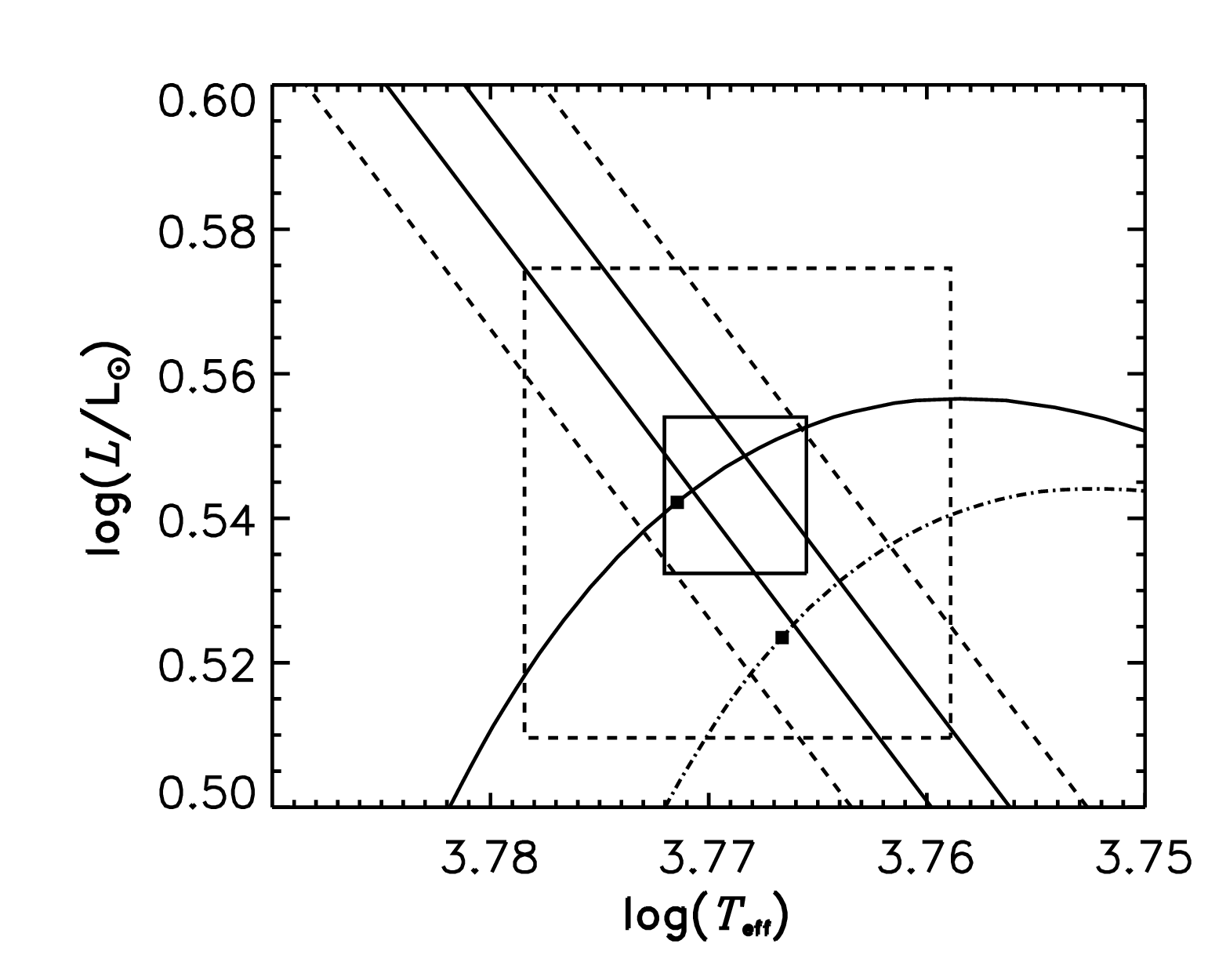}\\
\end{array}$
\end{center}
\caption{Left panel: The position of \bhyi~in the HR diagram. The constraints on the fundamental parameters ($T_{\rm{eff}}$, $L/\rm{L_{\odot}}$) are indicated
 by the 1-$\sigma$ error box (solid) and on the radius by diagonal
 solid lines. We also show the corresponding 3-$\sigma$ uncertainties
 by dashed lines. Two evolutionary tracks for the best models found
  by method 2 (cf. Table\,\ref{tab:res} ) are plotted with dash-dotted and solid
 curves, representing the models with and without gravitational
 settling and diffusion, respectively. Right panel: the same as in the left panel but zoomed in. The selected models are marked by filled squares.} \label{fig:hr}
\end{figure*}


\section{Observational constraints}
\label{sect:obscons}
\subsection{Non-seismic data}

The most recent determination of the
radius of \bhyi~is given by \cite{kjeldsen08}. The radius was
obtained by combining the interferometric angular diameter of the star,
$\theta=2.257 \pm 0.019$ mas \citep{north07}, with the revised
\emph{Hipparcos} parallax, $\pi_{\rm{p}}= 134.07 \pm 0.11$ mas \citep{van07}.

The luminosity, $L$, of a star can be obtained through the relation,
\begin{equation}
\label{eq:l}
L = 4 \pi F_{\rm bol} C^{2} / \pi_{\rm{p}}^{2},
\end{equation}
where $F_{\rm bol}$ is the bolometric flux and $C$ is the conversion
factor from parsecs to metres. To compute the luminosity
of \bhyi~we used the same value for the bolometric flux as \cite{north07},
$F_{\rm bol} = (2.019\pm 0.05)\times 10^{9}$ W m$^{-2}$
\citep[][the uncertainty on $F_{\rm bol}$ is from
\citealt{dibenedetto98}]{blackwell98} and the revised
\emph{Hipparcos} parallax \citep{van07}. Adopting L$_{\odot}=3.842\times 10^{26}$\,W
with an uncertainty of 0.4\% \citep{bahcall01}, we found
$L=3.494\pm 0.087$ L$_{\odot}$ for \bhyi.

A number of determinations of the
effective temperature ($T_{\rm{eff}}$) of \bhyi~
can be found in the literature
\citep[e.g.][]{favata97,blackwell98,dibenedetto98,santos05,dasilva06,biazzo07,bruntt10}.
We adopted the value of \cite{north07}, which
is derived from the direct measurement of the angular diameter.

The most recent value for the metallicity of \bhyi~is given
by \cite{bruntt10}, [Fe/H]=$-0.10\pm 0.07$. This value
is in agreement with the one found by \cite{santos05},
[Fe/H]=$-0.08\pm 0.04$ and with the one
adopted by \cite{fernandes03},
[Fe/H]=$-0.12\pm 0.047$. In our analysis we adopted
the metallicity from \cite{bruntt10}.

We calculated the mass fraction of metals, $Z$, from the
metallicity using the following approximation, valid for Population
I stars which do not present the $\alpha$-elements enrichment seen in metal
deficient stars \citep{wheeler89}:

\begin{equation}
\begin{split}
\label{eq:met}
\rm{[Fe/H]_s} & \equiv {\rm{log}} \left( \frac{Z_{\rm{Fe}}}{Z} \right)_{\rm{s}} + {\rm log} \left( \frac{Z}{X} \right)_{\rm{s}} - {\rm log} \left(\frac{Z_{\rm{Fe}}}{Z} \right)_{\odot} - {\rm log} \left( \frac{Z}{X} \right)_{\odot}\\
& = {\rm log} \left( \frac{Z}{X} \right)_{\rm{s}}- {\rm log} \left( \frac{Z}{X} \right)_{\odot},
\end{split}
\end{equation}
where $\rm{[Fe/H]_s}$ is the star's metallicity; $Z_{\rm{Fe}}$
and $X$ are the iron and hydrogen mass fractions, respectively; and
$(Z/X)_\odot$
is the ratio for the solar mixture. We used $(Z/X)_\odot$ = 0.0245
\citep{grevesse93}. This gives $(Z/X) = 0.019 \pm 0.003$ for \bhyi.

From spectral analysis, \cite{dravins90} found $v \sin i = 2 \pm 1$ km s$^{-1}$ for \bhyi. More recently,
\cite{bruntt10} found $v \sin i = 2.7 \pm 0.6$ km s$^{-1}$, and \cite{hekker10}
from spectroscopic line-profile analysis, found $v \sin i = 4.3$ km s$^{-1}$.
From their analysis, \cite{hekker10} attempted to determine the inclination angle, $i$, of \bhyi,
suggesting a value of $55 \pm 17^\circ$ for this star.
Thus the effect of rotation on the modelling of the structure of the star can be
neglected. Similarly, since the resulting rotational splitting is comparable with the
error in the observed frequencies (see below), in the present analysis we neglect the
effects of rotation on the frequencies.

\begin{table}[t]
\caption{Stellar properties of \bhyi. The luminosity, $L$, and
radius, $R$, are expressed in solar units. $\theta$ stands for the
angular diameter, $\pi_{\rm{p}}$ for the \emph{Hipparcos}
parallax, $T_{\rm{eff}}$ for the effective temperature,
[Fe/H] is the metallicity, and $Z/X$ is the mass ratio of heavy elements to hydrogen.}
 \begin{center}
\label{tab:param}
 \begin{tabular}{@{}lll}
 \hline
  & Value & Reference \\
 \hline
$\theta$ (mas) & 2.257 $\pm$ 0.019 & \citet{north07}
\\ 
$\pi_{\rm{p}}$ (mas)& $134.07 \pm 0.11$& \citet{van07}
\\
 $R/\rm{R_{\odot}}$ & 1.809 $\pm$ 0.015 & \citet{kjeldsen08}
 \\
 $L/\rm{L_{\odot}}$ & 3.494 $\pm$ 0.087
  &  Current work\\
 $T_{\rm{eff}}$ (K) & 5872 $\pm$ 44 & \citet{north07}
 \\
 $\rm{[Fe/H]}$&$-0.10 \pm 0.07$
  & \cite{bruntt10}
  \\
$Z/X$ & $0.019 \pm 0.003$&  Current work
\\
\hline
 \end{tabular}
 \end{center}
 \end{table}
The position of \bhyi~in the HR diagram is shown in Fig.\,\ref{fig:hr}
and the fundamental parameters we adopted are given in
Table\,\ref{tab:param}.
\subsection{Seismic data}
Asteroseismic observations of \bhyi~have been reported by
\cite{bedding07}. They found an excess power centred around 1 mHz with a
peak amplitude of $\sim$50 cm s$^{-1}$, and the oscillation
frequencies show a comb-like structure typical of solar-like
oscillations with a large frequency separation of the
$l = 0$ modes, $\Delta\nu_{0}$, of
57.24 $\pm$ 0.16 $\mu$Hz. They also identified 28 mode frequencies
in the range $0.7 < \nu < 1.4$ mHz with the angular degrees $l = $0,
1, and 2, three of which were identified as $l = 1$ mixed modes. In
this work, we used the updated list of 33 observed frequencies given in
Table\,\ref{tab:freqs}. To produce these, we reanalysed the 2005 dual-site
observations \citep{bedding07} using revised weights that were
adjusted using a new method that minimises the sidelobes (H.
Kjeldsen et al., in preparation).  This method is described by
\cite{bedding10}, who applied it to multi-site observations of
Procyon \citep[see also][]{arentoft09}. In the same way as for Procyon, we extracted
oscillation frequencies from the time series of $\beta$~Hyi using
the standard procedure of iterative sine-wave fitting.  The finite
mode lifetime causes many modes to be split into two or more peaks
which, coupled with the presence of mode bumping, meant that
deciding on a final list of mode frequencies with correct $l$
identifications is somewhat subjective. We followed the same
approach as \cite{bedding10}, which involved using the ridge
centroids as a guide and averaging multiple peaks into a single
value. The remaining unidentified peaks in the power spectrum are
listed in Table\,\ref{tab:noid}.
\begin{table}[h]
\begin{center}
\caption{Observed oscillation frequencies in \bhyi~ (in $\mu$Hz) resulting from the
revised analysis, listed in ascending radial order within each column. The rows are in
ascending $l$, and each row includes frequencies within $\Delta\nu$-sized-bits of the frequency
spectrum. "..." is used for the modes whose S/N was too low for a clear extraction.}
\scriptsize\begin{tabular}{@{}cccc} \hline \hline
$l=0$& $l=1$ & $l=2$ & $l=3$\\
\hline
660.74 $\pm$ 2.43&...&...& ... \\
716.68  $\pm$ 3.00&...&711.24 $\pm$ 2.13&...\\
774.79  $\pm$ 2.20&802.74  $\pm$ 1.69&769.97  $\pm$ 0.99&791.66  $\pm$ 1.35\\
831.86  $\pm$ 2.43&857.32  $\pm$ 0.86&825.86  $\pm$ 1.18&...\\
889.15  $\pm$ 1.23&912.91  $\pm$ 0.86& 883.35  $\pm$ 0.89&...\\
946.11  $\pm$ 0.91&959.98  $\pm$ 0.89&939.97  $\pm$ 0.97&...\\
...&987.08  $\pm$ 0.87& ...&...\\
1004.32  $\pm$ 0.86&1032.99  $\pm$ 0.86&999.40 $\pm$ 0.91&...\\
1061.66  $\pm$ 0.95&1089.87  $\pm$ 0.88&1057.00  $\pm$ 0.86&...\\
1118.67  $\pm$ 0.88&1147.35  $\pm$ 0.91&1115.20  $\pm$ 1.06&...\\
1177.76  $\pm$ 0.97&1203.54  $\pm$ 1.01&1172.98  $\pm$ 0.86&1198.16  $\pm$ 1.23\\
1235.31  $\pm$ 1.09&...&...&...\\
...&1320.42  $\pm$ 0.94&...&...\\
...&1378.92  $\pm$ 1.39&...&...\\
\hline
\end{tabular}
\label{tab:freqs}
\end{center}
\end{table}

\begin{table}[h]
\begin{center}
\caption{Unidentified observed peaks with S/N $\geq$ 3.5.}
\begin{tabular}{@{}ccc}
\hline \hline
\multicolumn{3}{c}{$\nu$ ($\mu$Hz)}\\
\hline
753.12  $\pm$ 1.57 & 1013.42  $\pm$ 1.50 & 1130.36  $\pm$ 1.30\\
828.70  $\pm$ 1.83 & 1025.80  $\pm$ 1.68 & 1134.32  $\pm$ 1.63\\
845.02  $\pm$ 1.61 & 1037.90  $\pm$ 1.63 & 1167.62  $\pm$ 1.10\\
868.60  $\pm$ 1.13 & 1065.12  $\pm$ 1.59 & 1256.78  $\pm$ 1.60\\
911.88  $\pm$ 1.76 & 1070.00  $\pm$ 1.43 & 1383.20  $\pm$ 1.75\\
1010.20  $\pm$ 1.91 & 1084.20 $\pm$ 1.57 & 1462.62  $\pm$ 1.92\\
\hline
\end{tabular}
\label{tab:noid}
\end{center}
\end{table}

\section{Modelling \boldmath \bhyi}
\label{sect:model}
\subsection{Input physics to the models}
To compute the evolutionary models we used the
\lq Aarhus STellar Evolution Code\rq, ASTEC \citep{cd08aastec}.
The following assumptions were made:
spherical symmetry, no rotation, no magnetic
field and no mass loss. In the computation we used the OPAL 2005
equation of state tables \citep[see][]{rogers02}, OPAL opacities
\citep{iglesias96} complemented by low-temperature opacities from \cite{ferguson05}, the
solar mixture from \cite{grevesse93}, and the NACRE nuclear reaction rates \citep{angulo99}.
We considered an atmospheric temperature versus optical depth
relation which is a fit to the quiet-sun relation of
\cite{vernazza76}.

Convection was treated
according to the standard mixing-length theory (MLT)
from \cite{vitense58}, where the characteristic length
of turbulence, $l_{\rm{mix}}$, scales directly with the local pressure scale height,
$H_{p}$, as $l_{\rm{mix}} = \alpha H_{p}$, leaving the scaling
factor $\alpha$ as a free parameter. MLT makes use of
another free parameter associated
with the amount of the core overshooting, $\alpha_{\rm{ov}}$.
Both \cite{dimauro03} and \cite{fernandes03} found that
models at the position of \bhyi~in the HR diagram are
not affected by convective overshooting, so we decided,
for this work,
not to consider it in our models.

Diffusion and settling were treated in the approximations proposed
by \cite{michaud93}. We refer to \cite{cd08aastec} for a
detailed explanation.

\subsection{Grids of models}
\begin{table*}[htbf]
\begin{center}
\footnotesize \caption{Parameters used to compute the evolutionary tracks. $M/\rm{M_{\odot}}$ is
the mass in solar units, $Z/X$ is the initial ratio of heavy elements to hydrogen abundances, and
$Y$ the helium abundance.}
\label{tab:grids}
\begin{tabular}{@{}lll}
\hline
Parameter & Grid I & Grid II\\
\hline
$M/\rm{M_{\odot}}$ & 1.00 - 1.18 (with steps of
0.02)&1.00 - 1.18 (with steps of 0.02)\\
$Z/X$ & 0.010 - 0.030 (with steps of 0.004)
&0.010 - 0.030 (with steps of 0.004)\\
$Y$ & 0.24 - 0.30 (with steps of 0.02)
 &0.24 - 0.30 (with steps of 0.02)\\
Mixing length & &\\
parameter ($\alpha$)& 1.4 - 2.0 (with steps of 0.2)
& 1.4 - 2.0 (with steps of 0.2)\\
Diffusion \& & & \\
{gravitational settling}& None & He \\
\hline
\end{tabular}
\end{center}
\end{table*}

We calculated two grids of evolutionary tracks, Grids I and II, with
the input parameters shown in Table\,\ref{tab:grids}. In Grid II we
included diffusion and gravitational settling of helium.

For each grid, we took those models whose parameters
were within the 3-$\sigma$ uncertainties derived from the observations
of \bhyi, and computed the corresponding
oscillation frequencies
using the Aarhus adiabatic oscillation code ADIPLS
\citep{cd08adipls}. The theoretical frequencies were
then compared with the observed ones in order
to find the best-matching model.

\section{Near-surface corrections}
\label{sect:corre} It is well known from helioseismology that there is a systematic offset
between the observed and the computed oscillation frequencies of the Sun. This offset, which is
nearly independent of the angular degree, $l$, of the mode and affects the highest frequencies
the most \citep{cd97}, arises from the improper modelling of the surface layers. Therefore, the
offset is also expected to be present when comparing the observed and computed frequencies for
other stars. \cite{kjeldsen08} used the solar data to derive an empirical correction for the
near-surface offset, which can, in principle, be applied to other stars.

For the Sun, the difference between the observed, $\nu_{\rm{obs}}$,
and computed frequencies of the best model, $\nu_{\rm{best}}$, was
shown by \cite{kjeldsen08} to be well approximated by a power law fit given as
\begin{equation}
\label{eq:corre1}
 \nu_{\rm{obs}}(n) - \nu_{\rm{best}}(n) = a \left[ \frac{\nu_{\rm{obs}}(n)}{\nu_{0}} \right]^b,
\end{equation}
where $n$ is the radial order, and $\nu_{0}$ is a reference frequency that is chosen to be
the frequency at maximum power. Since
the offset is independent of $l$, the authors considered
only radial ($l=0$) modes.
Note that the \lq best model\rq~is the one that best
represents the interior but still fails to
model the near-surface layers. They also argued that the
frequencies of a reference model, $\nu_{\rm{ref}}$, which is close to the
best one, can be scaled from $\nu_{\rm{best}}$ by a factor $r$, i.e.,
\begin{equation}
\label{eq:corre_r}
 \nu_{\rm{best}}(n) = r \nu_{\rm{ref}}(n).
\end{equation}
Then Eq.\,(\ref{eq:corre1}) becomes
\begin{equation}
\label{eq:corre2}
 \nu_{\rm{obs}}(n) - r \nu_{\rm{ref}}(n)  = a \left[ \frac{\nu_{\rm{obs}}(n)}{\nu_{0}} \right]^b.
\end{equation}
\cite{kjeldsen08}, using
the data and models for the Sun, found $b=4.90$. Using this
value for $b$ it is possible to determine $r$ and $a$ from
Eqs (6) and (10) in \cite{kjeldsen08}. Assuming that a similar offset occurs for other solar-like stars,
they showed how to use the solar $b$ to determine $r$ and $a$ from the observed frequencies of $l=0$ modes,
and, consequently, use Eq.\,(\ref{eq:corre2}) to calculate the correction that must be applied
to the frequencies computed for all $l$ for
a given stellar model. They noted that the correction applied to the mixed
modes should be less than that for the pure p modes (see the next section).

\section{Selecting the best model}
\label{sect:bestm} \label{bestm} We considered two methods to find the model that to the
closest extent possible reproduces the observed non-seismic and seismic data for \bhyi.
For all models whose parameters were within the  3-$\sigma$ uncertainties derived from the
non-seismic observations of \bhyi, we calculated the $r$ and $a$ values, following
\cite{kjeldsen08}, using $b=4.90$ and $\nu_{0}=1000\,\mu$Hz.

In the first method we followed closely the work of
\cite{kjeldsen08} and considered the best representative model to be the one
having the value of $r$ closest to 1,
which means, from Eq.\,(\ref{eq:corre_r}),
\begin{equation}
\label{eq:corre_r1}
 \nu_{\rm{ref}}(n) \thickapprox \nu_{\rm{best}}(n).
\end{equation}
Using the values of $r$ and $a$ found for this model, we then computed the correction factor to
be applied to the model frequencies and compared them with those observed. Note,
however, that in this method $r$ is calculated using only the observed radial modes together with
the corresponding model radial modes. If we assume that the best model is the one which has $r$
closest to 1 then we are assuming the model that has the radial frequencies matching most
closely the observed $l=0$ modes also has the nonradial frequencies matching best the
observed $l=1$ and 2 modes, which may not be true.

In the second method we performed a statistical analysis, using the $\chi^2$ test. Our goal was
to find the reference model that, after applying the surface corrections, had the individual
frequencies closest to the observed ones, for all $l$, i.e. we found the minimum of
\begin{equation}
\label{eq:chi2}
 \chi^2=\frac{1}{N} \sum_{n,l} \left(\frac{\nu_{\rm{ref,corr}}(n,l)-\nu_{\rm{obs}}(n,l)}{\sigma (\nu_{\rm{obs}}(n,l))}\right)^2,
\end{equation}
where $N$ is the total number of modes considered, $\nu_{\rm{ref,corr}}(n,l)$
are the frequencies of modes with radial order $n$ and degree $l$ of a reference
model, corrected for the surface effects, and $\sigma$ represents
the uncertainty in the observed frequencies.

The correction term, as shown in the right-hand side of Eq.\,(\ref{eq:corre1}) can
only be applied to the frequencies of the best model.
In order to compute the correction term, since we have
a set of reference models and we do not know which one is the best model,
we assume that the corrected best $\nu_{\rm{best,corr}}$ and reference model
$\nu_{\rm{ref,corr}}$ frequencies also scale as
\begin{equation}
\label{eq:corr3}
\nu_{\rm{best,corr}} (n,0) = r \nu_{\rm{ref,corr}}(n,0).
\end{equation}
We note that this is a good approximation because the surface corrections to the frequencies are
much smaller than the frequencies themselves. Moreover, since $\nu_{\rm{best,corr}} \simeq
\nu_{\rm{\rm{obs}}}$ we have
\begin{equation}
\label{eq:corr5}
\nu_{\rm{ref,corr}} (n,0) = \nu_{\rm{ref}}(n,0)+ \left(\frac{a}{r}\right) \left[\frac{\nu_{\rm{obs}} (n,0)}{\nu_0}\right]^b.
\end{equation}
Thus, Eq.\,(\ref{eq:chi2}) becomes
\begin{equation}
\label{eq:chi2m1}
\chi^2=\frac{1}{N} \sum_{n,l} \left(\frac{\nu_{\rm{ref}}(n,l)+\left(\frac{a}{r}\right) \left[\frac{\nu_{\rm{ref}} (n,l)}{\nu_0}\right]^b-\nu_{\rm{obs}}(n,l)}{\sigma (\nu_{\rm{obs}}(n,l))}\right)^2.
\end{equation}
Note that, in practice, the term $(a/r) [\nu_{\rm obs}(n,0)/\nu_0]^b$ on the right-hand
side of Eq.\,(\ref{eq:corr5}) was replaced by $(a/r) [\nu_{\rm ref}(n,l)/\nu_0]^b$. The reason is
to enable us to correct all the reference model frequencies, instead of only the frequencies
having the same radial order as those observed. The difference between these terms is
minimal for the best-fitting model, and so we are safe in performing the replacement.

To compute $\chi^2$ we used all the observed and computed $l=0,1$, and 2 frequencies. However, as
already mentioned, some of the observed $l=1$ modes are mixed modes and the scaling is not valid
for them. Mixed modes should not be affected by the surface layers as much as the p modes
\citep{kjeldsen08}, so the correction term should be small for the mixed modes. Specifically, at
a given frequency we expect the near-surface effects to scale inversely with the mode inertia,
which is much higher for the mixed modes than for the p modes; thus we scaled the correction term
by the inverse of the ratio $Q_{nl}$ between the inertia of the mode and the inertia of a radial
mode of the same frequency \citep{aerts10}. Taking that into account, Eq.\,(\ref{eq:chi2m1})
becomes
\begin{equation}
\label{eq:chi2m2}
\chi^2=\frac{1}{N} \sum_{n,l} \left(\frac{\nu_{\rm{ref}}(n,l)+\left(\frac{1}{Q_{nl}}\right) \left(\frac{a}{r}\right) \left[\frac{\nu_{\rm{ref}} (n,l)}{\nu_0}\right]^b-\nu_{\rm{obs}}(n,l)}{\sigma (\nu_{\rm{obs}}(n,l))}\right)^2.
\end{equation}
In this method, our best model is the one
that minimises Eq.\,(\ref{eq:chi2m2}).

Contrary to the first method,
the $\chi^2$ test
takes into account all the individual frequencies,
including mixed modes. This test also includes
the uncertainties in the observations. Thus, it is
possible that the model with the lowest $\chi^2$
does not have the $r$ that is closest to 1.

\section{Results and Discussion}
\label{sect:res_dis}
\begin{figure*}[ht]
\begin{minipage}[t]{0.5\linewidth} 
\centering
\includegraphics[width=8cm]{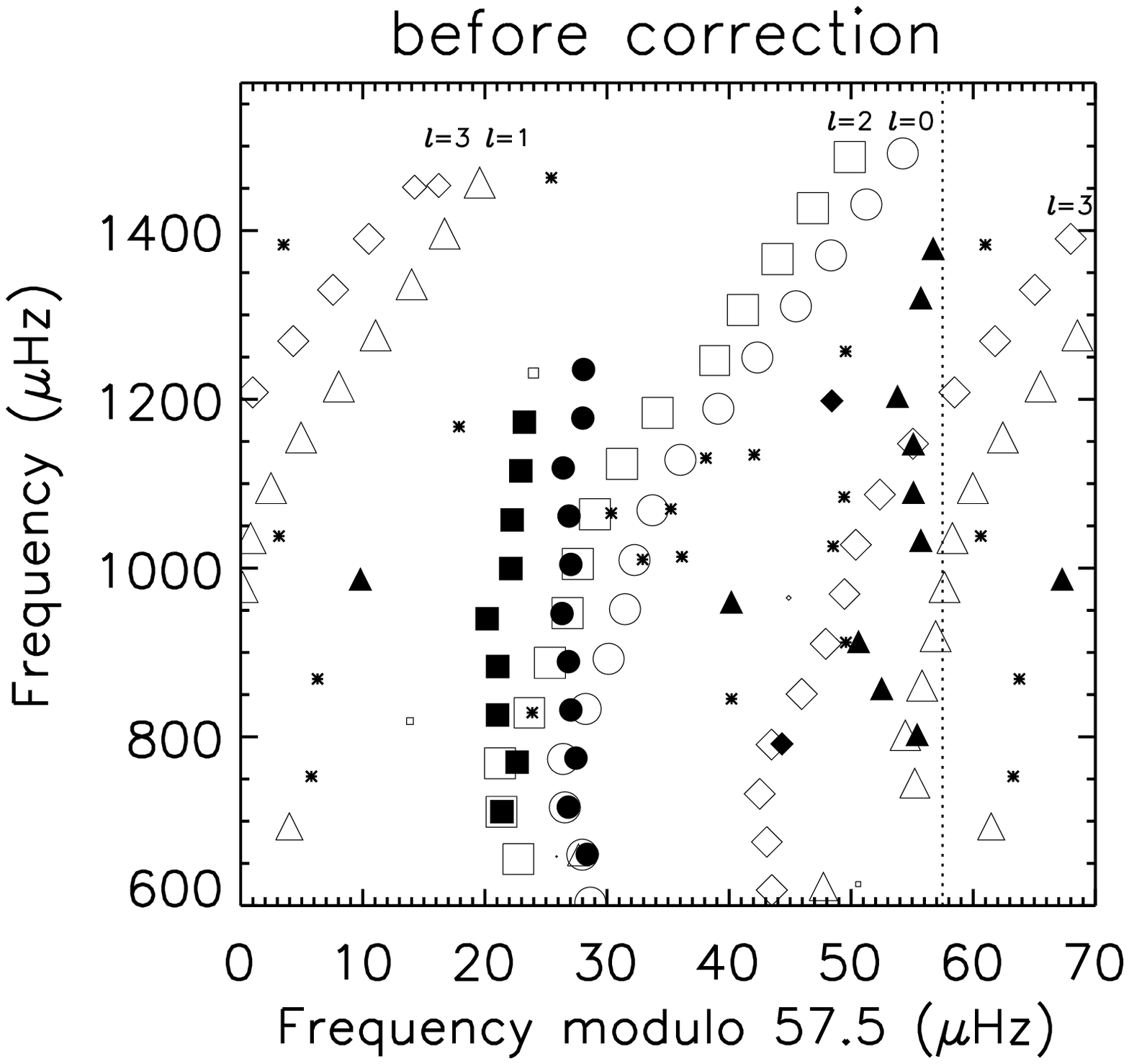}
\end{minipage}
\begin{minipage}[t]{0.5\linewidth}
\centering
\includegraphics[width=8cm]{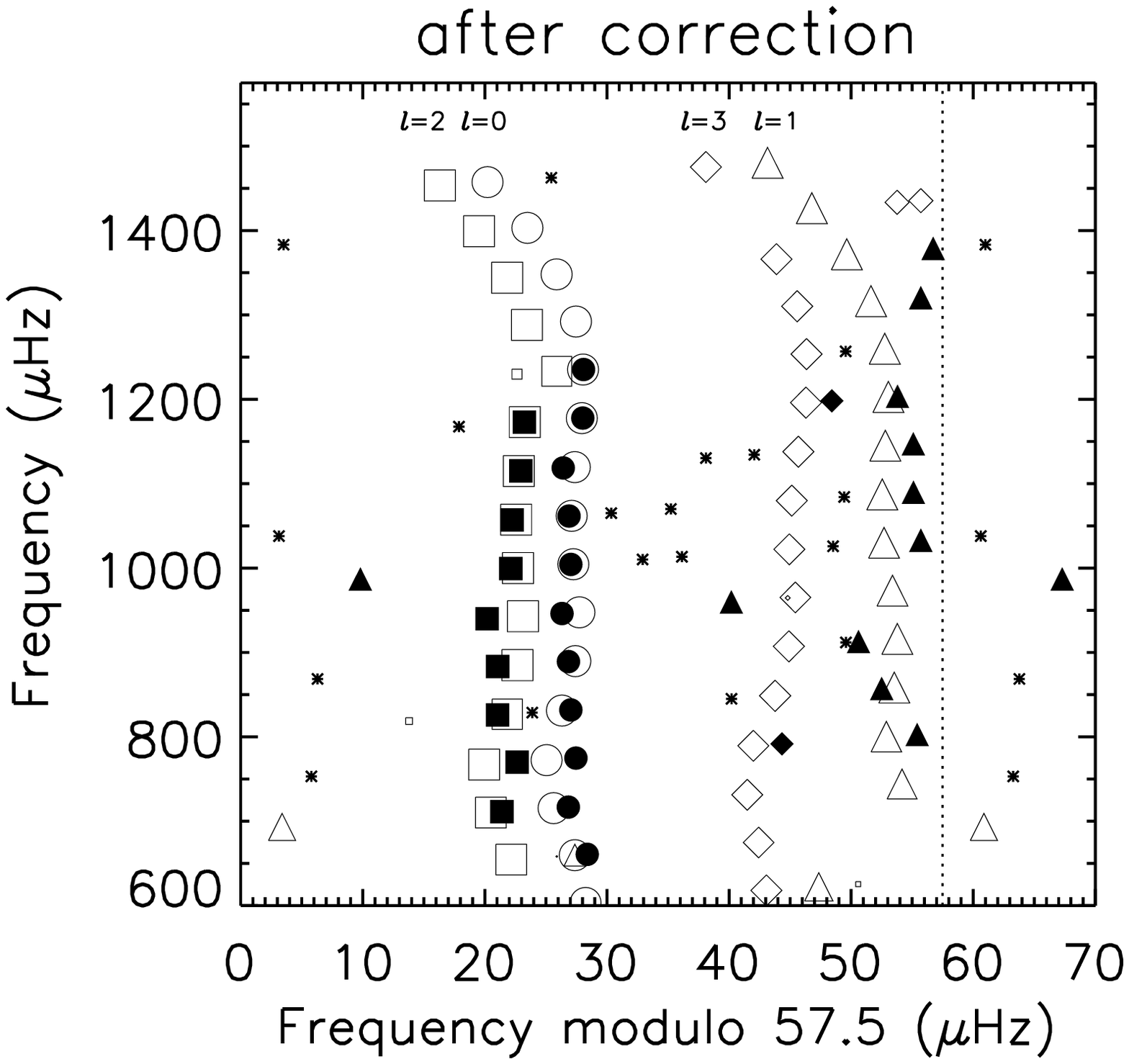}
\end{minipage}
\begin{minipage}[t]{0.5\linewidth} 
\centering
\includegraphics[width=8cm]{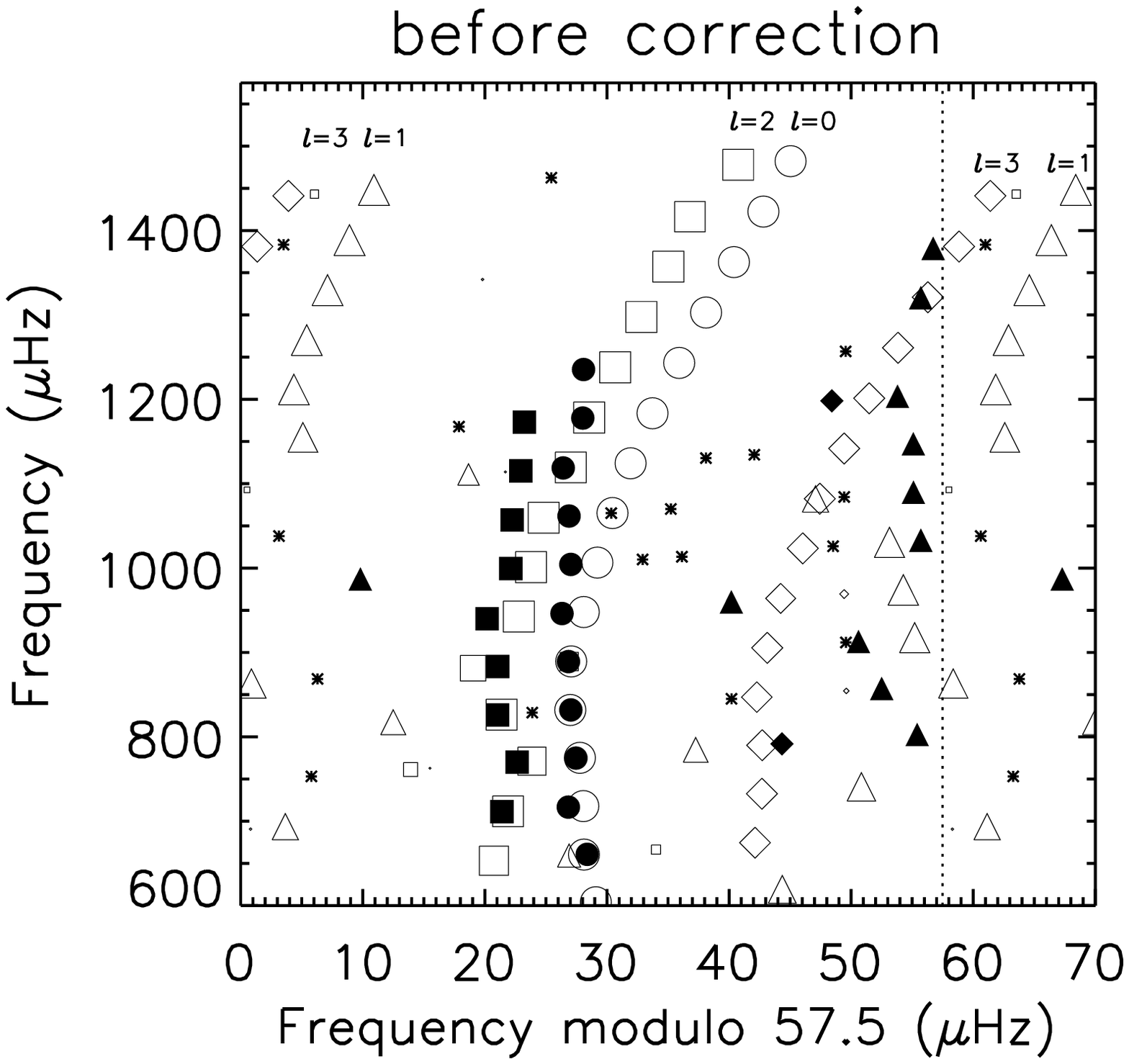}
\end{minipage}
\begin{minipage}[t]{0.5\linewidth}
\centering
\includegraphics[width=8cm]{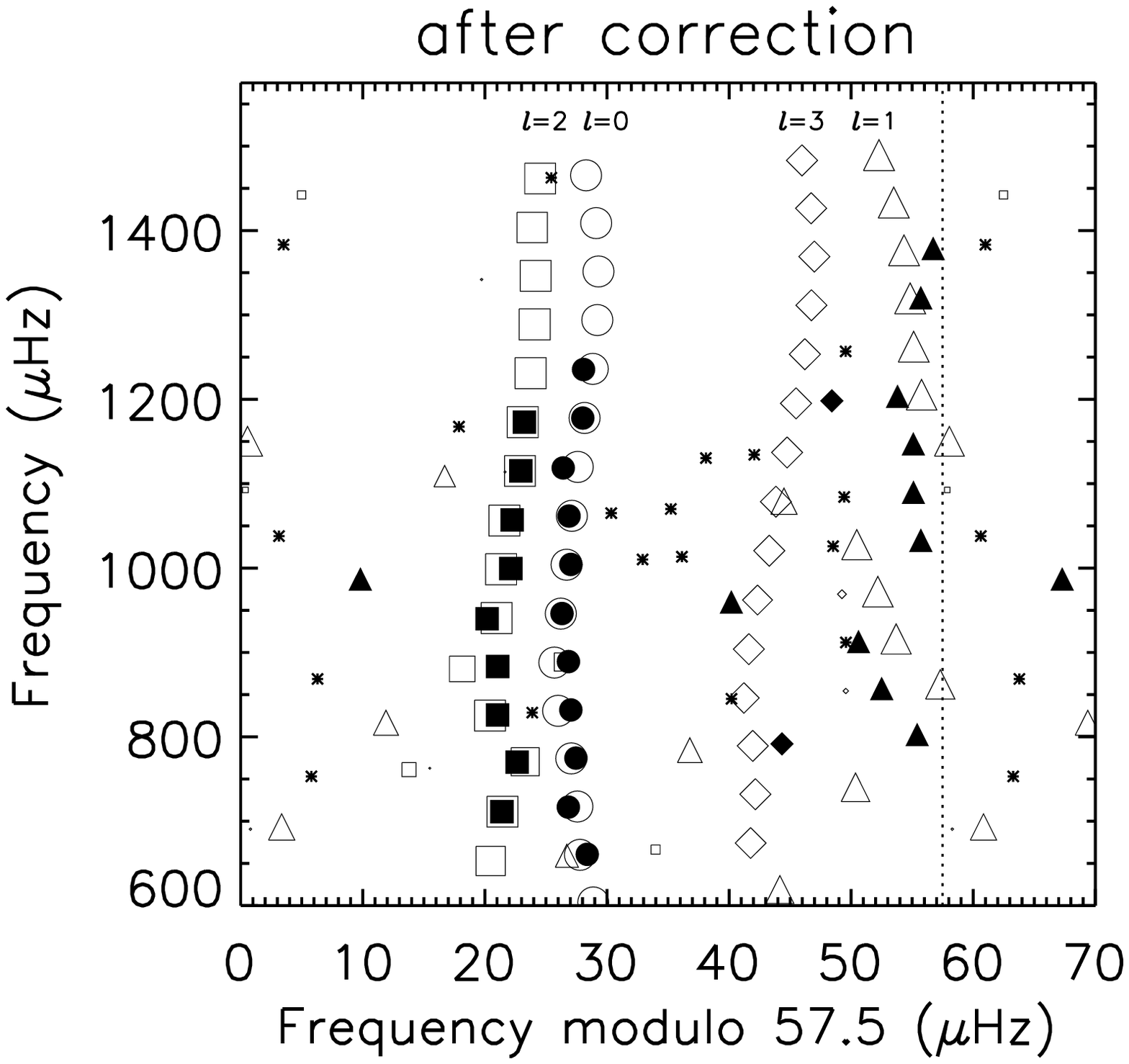}
\end{minipage}
\caption{\'Echelle diagrams for $\beta$\,Hyi, with a frequency separation of $\Delta \nu = 57.5\,\mu$Hz, 
before (left plot) and after (right plot) application of the near-surface corrections to the model frequencies. 
Shown are the frequencies of the selected models using method 1, when including no diffusion (upper panel)
 and diffusion (lower panel). In method 1, the best model was selected using the radial ($l=0$) modes alone 
(see the text for details). The solid symbols show observed frequencies (Table\,\ref{tab:freqs}), 
asterisks the unidentified peaks (Table\,\ref{tab:noid}), and the open symbols the model frequencies. 
Circles are used for $l=0$ modes, triangles for $l=1$, squares for $l=2$ and diamonds for $l=3$. 
Open symbols are scaled to represent the relative amplitudes of the modes as predicted by the models.}
\label{fig:resm1}
\end{figure*}

\begin{figure*}[ht]
\begin{minipage}[t]{0.5\linewidth} 
\centering
\includegraphics[width=8cm]{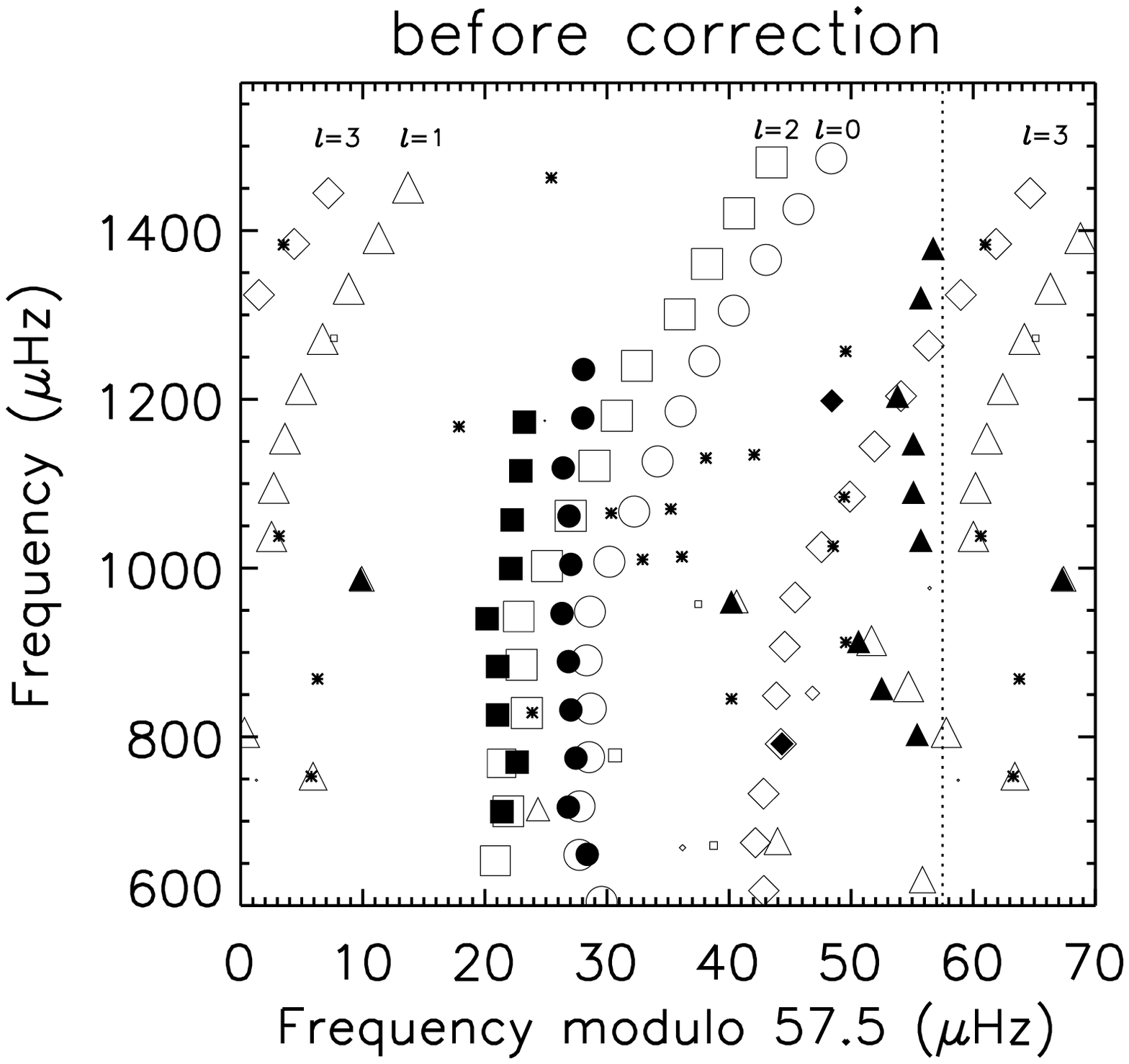}
\end{minipage}
\begin{minipage}[t]{0.5\linewidth}
\centering
\includegraphics[width=8cm]{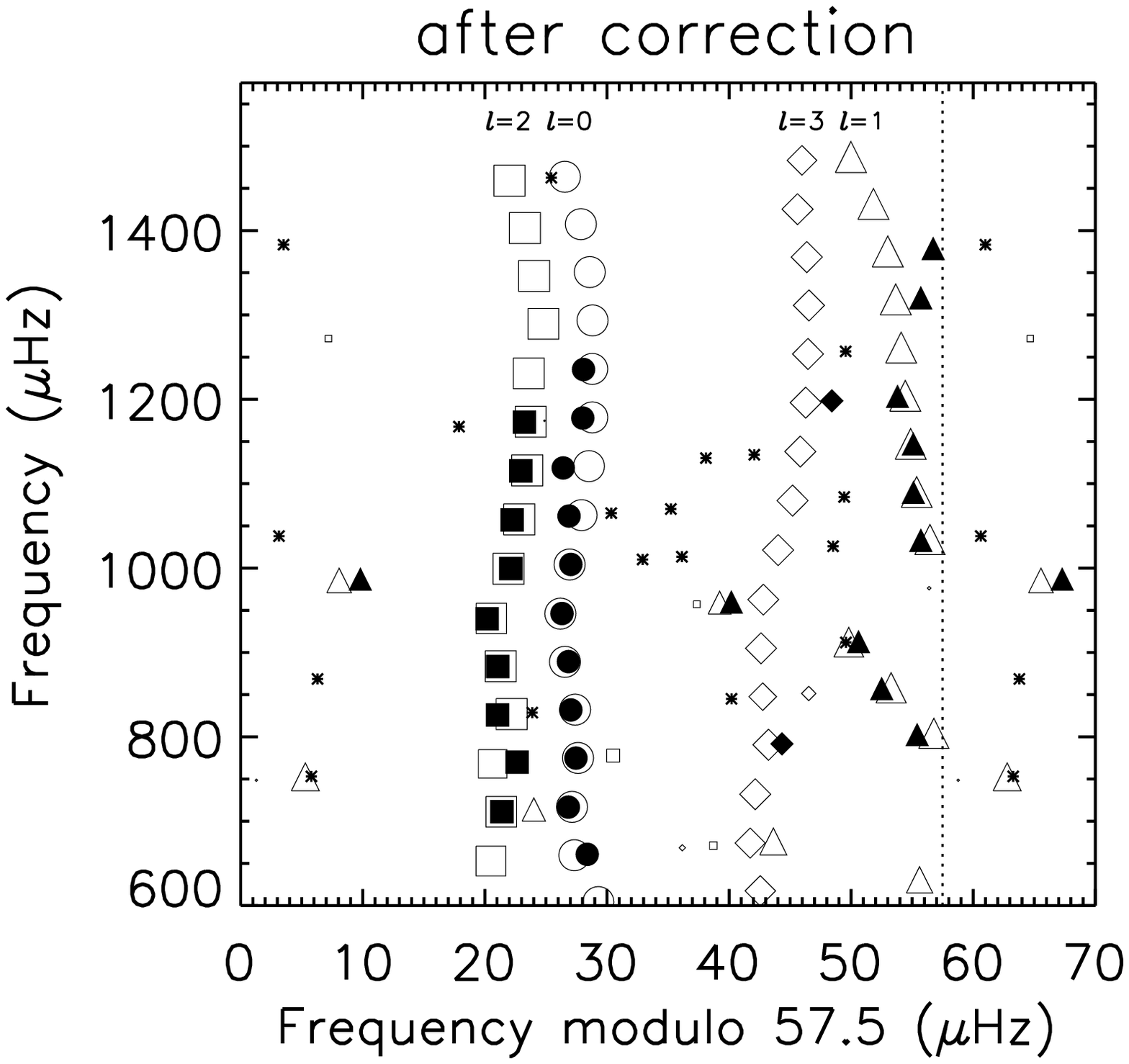}
\end{minipage}
\begin{minipage}[t]{0.5\linewidth} 
\centering
\includegraphics[width=8cm]{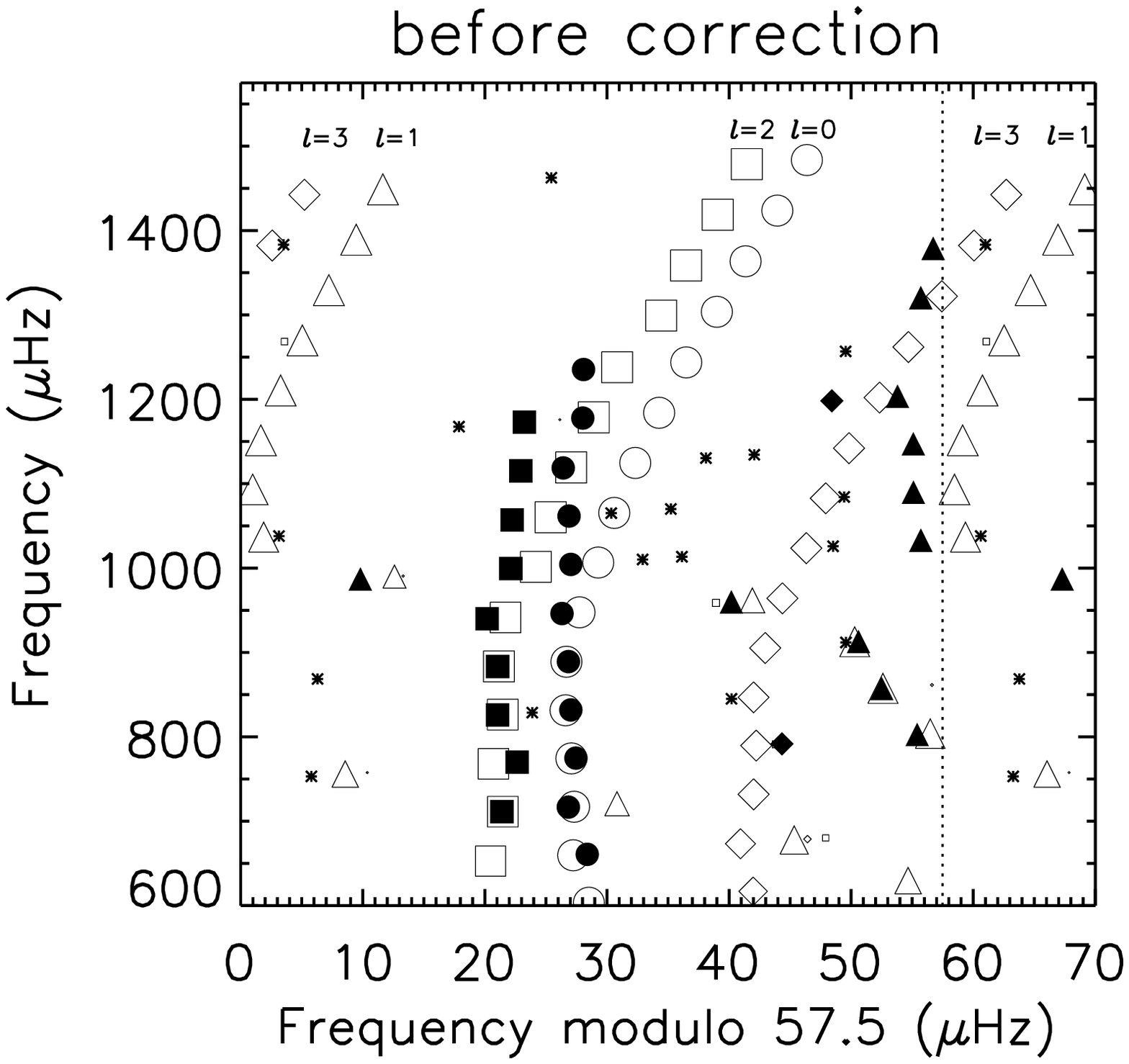}
\end{minipage}
\begin{minipage}[t]{0.5\linewidth}
\centering
\includegraphics[width=8cm]{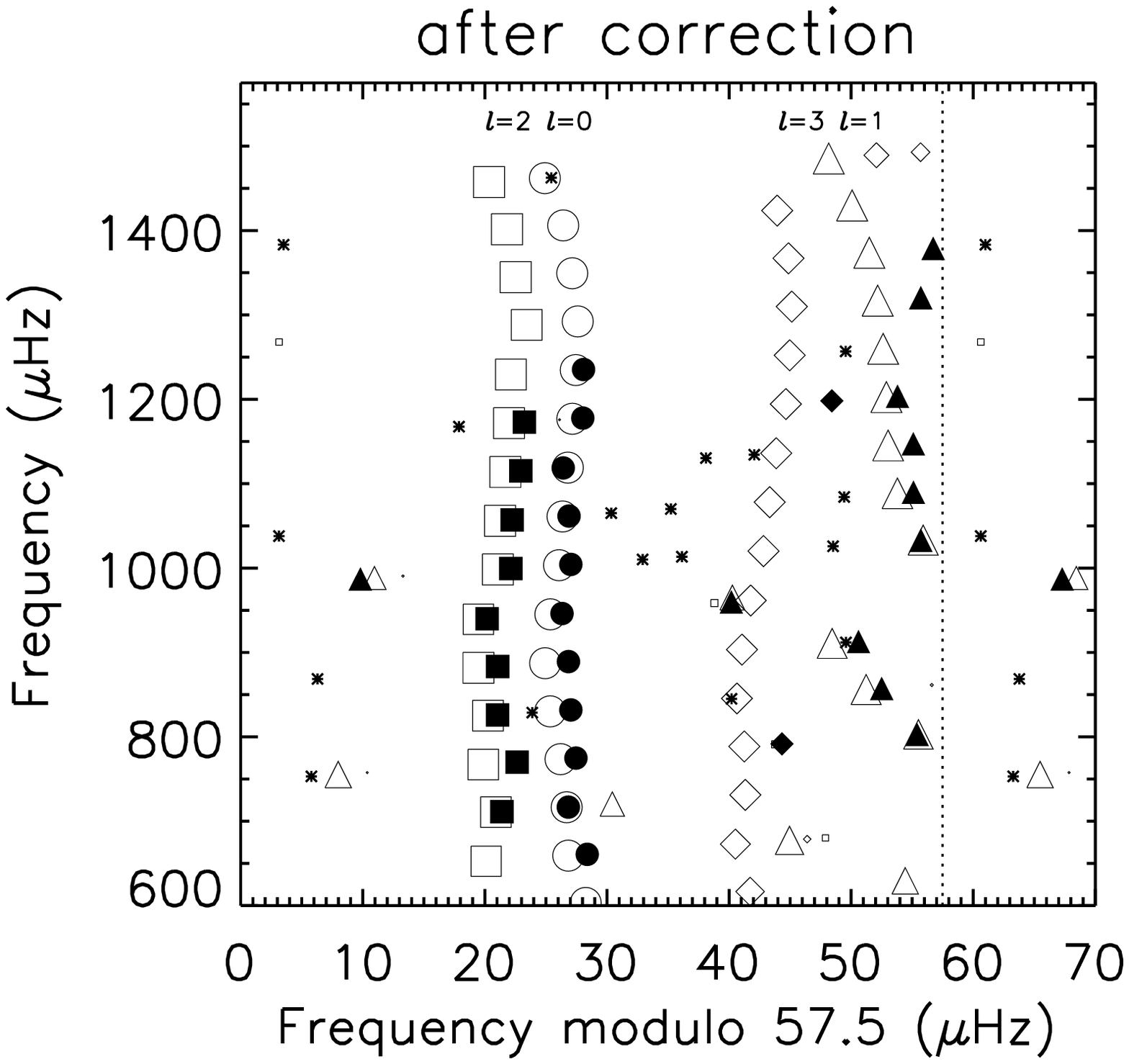}
\end{minipage}
\caption{The same as Fig.\,\ref{fig:resm1} but for the best models without (upper panel) and with (lower panel) diffusion, 
selected using method 2, which takes into account the observed and identified modes with all degrees available.}
\label{fig:resm2}
\end{figure*}

\begin{table*}[htbf]
\begin{center}
\centering \footnotesize \caption{The parameters of the best models found for Grid I (no
diffusion) and II (He settling and diffusion), for each of the two methods. See text for details
on the methods. The mass, $M$, luminosity, $L$, and radius, $R$, are expressed in solar units.
$T_{\rm{eff}}$ is the effective temperature, $Y$ and $Z$ are the initial helium and
heavy-element abundances, [Fe/H] is the metallicity at the surface, and $\alpha$ is the
mixing-length parameter. Also $r$ and $a$ are factors used to compute the
correction term, $\Delta \nu_{0b}$ and $\Delta \nu_{0a}$ are, respectively, the large frequency
separation before and after applying the surface correction to the model $l=0$ modes. The
values of $\chi^2$ are those calculated after correcting the frequencies for the near-surface
effects.} \label{tab:res}
\begin{tabular}{l|ll|ll} \hline \hline
   & \multicolumn{2}{c|}{Grid I} & \multicolumn{2}{c}{Grid II} \\ \hline
 Parameter &  Method 1    &  Method 2  & Method 1&  Method 2  \\
\hline
$M/\rm{M_{\odot}}$ &1.16 & 1.04&1.04&1.04
\\
$R/\rm{R_{\odot}}$ &1.832 &1.785  &1.790&1.786
\\
$L/\rm{L_{\odot}}$ &3.433 & 3.485&3.432&3.338
\\
$T_{\rm{eff}}$ (K) & 5810&5908 &5877&5843
\\
Age (Gyr)&4.705& 6.114&7.390&7.297
\\
$Z$& 0.0204& 0.0124& 0.0075&0.0075
\\
$Y$& 0.30& 0.30&0.24&0.24
\\
$[$Fe/H$]$& 0.088& -0.133 & -0.416 & -0.424
\\
$\alpha$&1.4& 1.8&2.0&1.8
\\
$r$& 1.0000& 0.9995& 1.0000&1.0009
\\
$a$ ($\mu$Hz)& -4.80 &-3.14 &-2.43&-3.11
\\
$\Delta \nu_{0b}$ ($\mu$Hz)& 58.977& 58.488&58.243&58.400
\\
$\Delta \nu_{0a}$($\mu$Hz)& 57.678& 57.652&57.600&57.577
\\
$\chi^2$ &19.086&1.183&26.226&2.642
\\
\hline
\end{tabular}
\end{center}
\end{table*}
The parameters of the best models found for Grids I and II, using
the two methods described in  Sec. \ref{bestm}, are shown in
Table\,\ref{tab:res}. Figures \ref{fig:resm1} and \ref{fig:resm2}  show the
\'echelle diagrams for \bhyi. An \textit{\'echelle} diagram shows
the frequency spectrum divided into segments equally spaced by the
large frequency separation, after these segments are stacked in
the vertical direction \citep{grec83}. In these figures the observed
frequencies of \bhyi~are compared with the theoretical frequencies
of the best models from Grid I (upper panel) and from Grid II (lower
panel), both before (left plot) and after (right plot), applying the
near-surface corrections. The model frequencies are represented by
open symbols and the observed frequencies (cf. Table\,\ref{tab:freqs}) by solid
symbols, while the asterisks represent the unidentified peaks
(cf. Table\,\ref{tab:noid}), which may correspond to genuine modes, sidelobes, or noise
peaks. The relative sizes of the open symbols reflect the expected
mode amplitudes \citep{jcd95}. The so-called mixed modes reveal themselves in the
\'echelle diagrams, breaking the regularity of the ridges. The
models predict mixed modes with all nonradial degrees, however
mostly with too small amplitudes to be observed. On the other hand,
some of the observed modes match well the mixed modes with $l = 1$ (see,
e.g., the right panels of Fig.\,\ref{fig:resm2}.
If we inspect Figs.\,\ref{fig:resm1} and \ref{fig:resm2}, it is
clear that the agreement between the observed and model frequencies
is much better when method 2 is used (Fig.\,\ref{fig:resm2}). This is due to the
fact that in this method all the available seismic constraints were
involved in selecting the best model.


It is seen from Table\,\ref{tab:res} that the model that has the $r$ value closest to unity does
not produce the lowest $\chi^2$ value. The model with the lowest $\chi^2$ still has an $r$
satisfactorily close to unity. So, in addition to finding a model that represents the stellar
interior reasonably, method 2 makes sure that all the available seismic constraints are
simultaneously reproduced and so the fit, and hence the accuracy of the model, is
improved substantially. This shows the importance of the individual modes in constraining the
range of models to represent the observed star. Mixed modes, in particular, put strong
constraints on the model properties, especially on the evolutionary stage. For instance, we can
see from the right panels of Fig.\,\ref{fig:resm1} that two models resulting from method 1 have
the two highest $\chi^2$ values due to failing to match particularly the observed $l=1$ mixed
modes. In the upper right panel, the model is too massive and it matches the rest of the seismic
constraints before it is sufficiently evolved to have mixed modes, whereas the model in
the lower right panel does have mixed modes, although the predicted mixed modes
do not match those observed.

In general, we found that the empirical surface corrections proposed
by \cite{kjeldsen08} work very well for \bhyi~as seen
from Figs.\,\ref{fig:resm1} and \ref{fig:resm2}, although there is room
for improvement for high-frequency modes of $l=1$. The reason for the
suboptimal agreement for those modes is that the correction term is
determined using only the $l=0$ observed modes, whose frequencies
span a smaller range than those of the $l=1$ modes. Thus, radial
modes with higher frequencies need to be detected in order to
improve the agreement for the higher frequency $l=1$ modes.
Note that the change in the large frequency separation of the models
after applying the near-surface correction is around 0.8 $\mu$Hz,
which is larger than the given uncertainty of the observed large
separation. This should be taken into account when modelling through a
pipeline analysis that uses the large separation as input.
It is encouraging to see that we can observe $l=3$ modes, and that some of
the unidentified modes are also close to the model frequencies,
namely $753.1\,\mu {\rm Hz}$ ($l$ = 1?) and $1462.6\,\mu {\rm Hz}$
($l$ = 0?).

Even though the best model seems to be the one without diffusion, we do
expect that within a star diffusion occurs. The two selected models, the
ones with and without diffusion, resulting from the method 2 are in
fact compatible and both could be further fine-tuned.

\section{Summary and Conclusions}
\label{sect:end} We computed two grids of evolutionary models, using the ASTEC code, in order to
find the model that most closely reproduces the seismic and non-seismic data of \bhyi.
The parameters used for each grid are given in Table\,\ref{tab:param}. We computed the
oscillation frequencies for the models that lie inside the 3-$\sigma$ error box of the
star's position in the HR diagram with the ADIPLS code, and compared them with the
observed frequencies. There is an offset needed to be taken into account in this
comparison, due to improper modelling of the near-surface layers of the stars. We used the
approach proposed by \cite{kjeldsen08} to correct the computed frequencies of \bhyi~from this
offset. We used two methods in order to find the model that reproduces the observed oscillation
frequencies of \bhyi.

In our analysis, we argue that the method involving the $\chi^2$ test,
method 2, is the most robust way to find the best
model, since it takes into account all the individual frequencies,
the mixed modes,
and also the uncertainties on the observed frequencies.
Analysing the \'echelle diagrams of the representative models found with
method 2 (cf. Sect.\,\ref{sect:res_dis}), we see that the surface correction works
very well for $l=0$ modes, and for $l=1$ and 2 modes with
frequencies lying in the frequency range of the observed radial
modes. This was expected since the correction term was computed
using only those radial modes. Observed $l=0$ modes with higher
frequencies are thus needed in order to improve the surface
correction.

Our best models give $M$=1.04 M$_\odot$ and an age of 6.1 -- 7.3 Gyr
for \bhyi, depending on the inclusion of gravitational settling and
diffusion of helium. In either case, the radius is found to be
$R\sim1.785\,\rm{R}_{\odot}$, which is in good agreement with the
one determined by interferometry, $R = 1.809 \pm 0.015$ R$_{\odot}$.
However, there are other models fitting the data similarly well. We
used the parameters of those models (with $\chi^2 < 10$) to
determine the internal error regarding our analysis. We calculated
the mean value, and the uncertainties were taken as the standard
deviation. We found $M=1.08\pm0.03$ M$_\odot$, age = $6.40 \pm
0.56$ Gyr, and $R = 1.811 \pm 0.020 $ R$_{\odot}$. These results
are also consistent with the results of \cite{fernandes03}, who
derived, $M=1.10_{-0.07}^{+0.04}$ M$_\odot$ and $M=1.09\pm 0.22$
M$_\odot$, through the HR diagram analysis and $\Delta\nu_0$,
respectively, and a stellar age between 6.4 and 7.1 Gyr.

\begin{acknowledgements}
This work was partially supported by the project
PTDC/CTE-AST/098754/2008 and the grant SFRH / BD / 41213 /2007 funded
by FCT / MCTES, Portugal. MC is supported by a Ci\^encia 2007
contract, funded by FCT/MCTES (Portugal) and POPH/FSE (EC). GD, JCD, and
HK are grateful for the financial
support from the Danish Natural Science Research Council. TRB is
supported by Australian Research Council. NCAR is supported by the U.S. National
Science Foundation.
\end{acknowledgements}
\bibliographystyle{aa}
\bibliography{isa_bhyi_ver10.bib} 

\end{document}